\documentclass[preprint,12pt]{elsarticle}

\usepackage{booktabs}
\usepackage[mathscr]{eucal}

\usepackage{graphicx}

\usepackage{amssymb}
\usepackage{amsmath}

\journal{Nuclear Instruments and Methods A}
\begin{document}
\begin{frontmatter}
\title{Isochronicity Correction in the CR Storage Ring}

\author[1]{S.~Litvinov}
\author[2]{D.~Toprek}
\author[1]{H.~Weick}
\author[1]{A.~Dolinskii}
\address[1]{GSI, Helmholtzzentrum f\"ur Schwerionenforschung GmbH, 64291 Darmstadt, Germany}
\address[2]{VINCA Institute of Nuclear Sciences, Belgrade University, P.O Box 522, 11000 Belgrade, Serbia}
%
\begin{abstract}
A challenge for nuclear physics is to measure masses of exotic nuclei up to the limits 
of nuclear existence which are characterized by low production cross sections and short 
half-lives.
The large acceptance Collector Ring (CR)~\cite{cr} at FAIR~\cite{fair} tuned in the 
isochronous ion-optical mode offers unique possibilities for measuring short-lived and 
very exotic nuclides.
However, in a ring designed for maximal acceptance, many factors limit the resolution. 
One point is a limit in time resolution inversely proportional to the transverse emittance.
But most of the time aberrations can be corrected and others become small 
for large number of turns.
We show the relations of the time correction to the corresponding transverse focusing
and that the main correction for large emittance corresponds directly to the 
chromaticity correction for transverse focusing of the beam.
With the help of Monte-Carlo simulations for the full acceptance
we demonstrate how to correct the revolution times so that in principle
resolutions of $\Delta m/m=10^{-6}$ can be achieved.
In these calculations the influence of magnet inhomogeneities and extended fringe fields are 
considered and a calibration scheme also for ions with different mass-to-charge ratio is 
presented.
\end{abstract}
\begin{keyword}
CR, achromat, storage ring, nuclear masses
\PACS codes 29.20.db, 21.10.Dr, 29.30.Aj

\end{keyword}

\end{frontmatter}


\section{Introduction}
The Collector Ring of the FAIR facility is a symmetric, achromatic ring with two arcs, two
straight sections and a total circumference of 221.5 meters. 
It is designed for operation at a maximum magnetic rigidity of $B\rho=13\,$Tm \cite{cr}.
It will be operated in three ion-optical modes,
two of them providing fast pre-cooling of either antiprotons or radioactive ion beams~\cite{cr}.
In the third mode (isochronous optics) the CR will be operated as a Time-of-Flight (ToF) 
spectrometer for the mass measurement of exotic very short-lived nuclei (T$_{1/2}>20 \mu{s}$)
produced and selected in flight with the Super-FRS fragment
separator~\cite{sfrs}. This technique for mass measurements has been
developed at the ESR at GSI~\cite{Hau2000} and meanwhile was also used successfully in 
the CSRe at IMS Lanzhou \cite{Tu2011} and the RIKEN Rare-RI ring is under construction \cite{Yam2008}.

The first basic condition for isochronicity is the well known equation \cite{Hau2000} 
which describes the 
revolution time ($T$) of ions stored in the ring and its deviation ($\Delta T$) depending on
the difference in mass-to-charge ratio ($m/q$) or different velocity ($\upsilon$):
\begin{equation} \label{eq1}
\frac{\Delta T}{T} = \frac{1}{\gamma^{2}_{t}} \cdot \frac{\Delta(m/q)}{(m/q)}
+\Big(\frac{\gamma^{2}}{\gamma^{2}_{t}}-1\Big)\frac{\Delta v}{v}
+\frac{\Delta T_{others}}{T},
\end{equation}
where $\gamma$ is the relativistic Lorentz factor and $\gamma_{t}$
is the transition energy of the ring. The isochronous condition is
reached when $\gamma$ = $\gamma_{t}$. This means, the second term
in Eq.~(\ref{eq1}) vanishes and $\Delta(m/q)$ can be determined from the observed $\Delta T$.
It is remarkable that $T$ is directly proportional to $m/q$. 
But in addition to the intended time differences for different masses and
the important first order isochronicity condition
the resolution depends on more additional influences 
on time-of-flight which are collected in $\Delta T_{others}$.
These aberrations can also have higher order contributions.
Effects of nonlinear field errors, fringe fields of magnets, closed orbit distortion
and the transverse emittance negatively act on the resolution~\cite{Dol2007}. 

The general layout of the CR has been described in ref.\cite{Dol2008}.
The isochronous mode has been calculated for different values of $\gamma_t$ \cite{stori11}. 
In this paper we will always refer to the mode for $\gamma_t=1.67$ which allows to measure masses 
up to $m/q=3.1\,u/e$. In this case the momentum acceptance becomes $\Delta p/p = \pm 0.5\%$ 
when having at the same time a transverse acceptance of $100\,$mm mrad in both planes.

A correction of time-of-flight aberrations due to a large transverse phase space 
distribution is essential or otherwise the goal of large phase acceptance in the 
CR cannot be achieved with sufficient time resolution \cite{Dol2007}.
The simple reason for this influence is that ions which oscillate around the optical axis 
have a longer path length for one turn which leads to a corresponding time spread.
As it was indicated before \cite{Wol90} and will be also shown in this article these 
time-of-flight aberrations are directly related to the chromatic aberrations of the 
optical system.
It is known that in a straight system with only magnetic lenses defined by magnets outside 
of the beamline a correction is not possible \cite{Cou71}. With additional electrostatic elements 
this would be possible \cite{Haw09}, however, these are not available for the high rigidity of the 
CR beam. At least a correction in a system with a bent optical axis is possible \cite{Haw09}.
A recipe for such a full second or higher order achromat was given by Brown \cite{Bro79},
and partially this was also applied in the spectrometer TOFI \cite{Wou85} for time-of-flight 
mass measurements of nuclear masses. However, when a storage ring is constructed in this way 
a four fold  symmetry system which is imaging violates the basic stability criterion of a 
nonzero phase advance for a closed ring \cite{Wol87b}. 

This is also the result of a general approach to higher order achromats by Wan \cite{Wan96}, 
the minimum symmetry needed for a second order achromat is a four-fold symmetry, but based on 
the Lie algebra description of the system \cite{Dra87} it is found that the tunes of 
the individual cells should be integer.
A good approximation to a second order achromat may be found with the method of extending a full 
achromat by straight matching sections for adjusting the tunes. In this case the overall second 
order geometric aberrations can be kept zero as pure quadrupoles are free of them and by an 
overcorrection with the sextupoles at least the tune shift be corrected (so-called pseudo achromat)
\cite{Ser87}. As the matching section can be short compared to the whole arc section this usually 
leads to systems close to a full achromat.

This dilemma may not be so crucial in a ring for only few revolutions but for commissioning and other 
detection techniques like Schottky pickups stability of motion must be guaranteed. In this paper we 
will show a solution that does not lead to full achromaticity over one turn but to a correction in 
the limit of many turns. By averaging the time of flight over many turns still a good resolution 
for a large phase space can be obtained.
This less stringent requirement also helps to avoid the usual problem of large higher order field 
components, which in turn could impose other higher order aberrations.

In the CR we want to use ToF detectors inside the ring which do measure the revolution time for each turn.
This means we must observe the ions over a series of turns and see how large the remaining error will be.
This is anyway necessary as the resolution of the detector itself is limited and only averaging over a 
larger number of turns helps to obtain the desired resolution.


%
\section{Calculation of Second-Order Time Deviations}

The relative revolution time difference between an arbitrary and the reference particle can be expressed in 
terms of the initial coordinates as a Taylor series in a second-order approximation~\cite{carey,yavor}:
\begin{multline}
\label{second-matrix-terms}
\frac{\Delta T}{T} = \frac{T - T_{0}}{T_{0}} = (t|x)_{c}x+(t|a)_{c}a+(t|\delta)_{c}\delta\,+(t|xx)_{c}x^{2}\\
+(t|xa)_{c}xa+(t|aa)_{c}a^{2}+(t|yy)_{c}y^{2}+(t|yb)_{c}yb\\
+(t|bb)_{c}b^{2}+(t|x\delta)_{c}x\delta+(t|a\delta)_{c}a\delta+(t|\delta\delta)_{c}\delta^{2},
\end{multline}
where ($x$),($y$) are the transverse position coordinates and ($a$),($b$) the transverse momenta 
divided by the forward momentum of a reference particle ($p_0$).
The coefficients of the Taylor series correspond to the partial derivatives of the first 
coordinate before the line with respect to the following.
The index $c$ marks coefficients normalized by the total time-of-flight $n\,T_{0}$, 
where n is the number of turns and $T_{0}$ is the revolution time of the reference particle.
The fractional momentum deviation $\delta$ is given by $p~=~p_{0}(1+\delta)$.

In the first-order achromatic ring, with $(x|\delta)=(a|\delta)=0$, the first-order transverse 
time coefficients $(t|x)$ and $(t|a)$ vanish simultaneously~\cite{Wol90}.
The necessary condition to be an isochronous ring in first order is
$(t|\delta)\, =\, 0$, i.e. $\gamma$ = $\gamma_{t}$.
The second-order isochronous condition is fulfilled when $(t|\delta\delta)\,=\, 0$, which
can be corrected with one family of sextupole magnets installed in a dispersive section of the ring.

With such a good adjustment of $\gamma_t$ even to higher order in an achromatic ring
the largest contribution to $\Delta T$ comes from the second-order geometric time aberrations. 
For a beam of one 
species in an ideal isochronous ring without higher order field errors or closed orbit 
distortions only pure betatron motion exists. In such a ring the time spread
is directly related to the transverse emittances ($\varepsilon_{x,y}$)~\cite{Dol2007}:
\begin{equation}
\label{freq_spread}
\Big(\frac{\Delta T}{T}\Big)_{emitt.}\approx\frac{1}{4}\left(\varepsilon_{x}<\gamma_{x}>
+\varepsilon_{y}<\gamma_{y}>\right),
\end{equation}
where
$\varepsilon_{x,y}$ is the transverse beam emittance, which can be regarded as the ring acceptance, 
and $<\gamma_{x,y}>$ are the Twiss parameters $\gamma_{x,y}$ averaged over the whole 
circumference of the ring. 
This result simply describes the longer path length of ions traveling with larger betatron
amplitude over one turn.

From the maximum deviation given in Eqs.~\ref{eq1},~\ref{freq_spread} one can derive the 
mass resolution over the whole beam emittance.
For the CR, with acceptance $100\,$mm mrad in both planes,
the limit of mass resolution would be about $10^{-5}$,
which is insufficient for precise mass measurements.
Therefore, in order to reach the necessary resolution of $10^{-6}$
the transverse emittance would have to be limited to $10$ mm mrad in both planes.
As a result, the transmission of the ions into the ring would be reduced drastically.
However, the mass resolving power can be improved using second-order corrections and keeping the 
transverse emittance large.

\begin{figure}[hb]
\centering
\includegraphics[width=93mm]{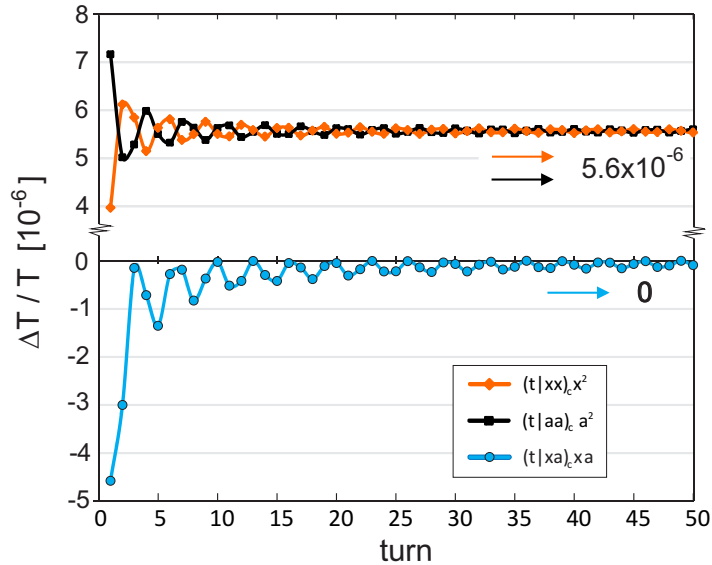}
\caption{Evolution of the relative second-order geometric time aberrations as a function 
of the number of turns in the CR for a single ion with start positions $x$ and $a$ on the edge of the 
acceptance of $100\,$mm mrad. The inserted numbers show the limit over many turns.}
\label{matrix_terms}
\end{figure}

For the second order terms it is useful to investigate their behavior over many revolutions.
As the system is bound and the betatron oscillations have a non-zero phase advance
the initial coordinates of a single ion change from turn to turn within fixed limits.
The time deviations over many turns simply sum up. The fluctuations are periodic with the 
tune and have an average value of zero. 
Therefore, the time deviations given by a single matrix coefficients which depend on mixed 
initial coordinates will oscillate with a fixed maximum amplitude. 
When looking at the relative time difference over many turns these contributions will become small.
The quadratic terms however will continue to increase from turn to turn and the relative time deviation
converges towards a constant value. Fig.\ref{matrix_terms} illustrates this behavior.

The constant second-order geometric aberrations in time of Eq.\ref{second-matrix-terms} are most significant.
Without correction their contribution corresponds to the right part of Eq.\ref{freq_spread} proportional 
to the transverse emittance.

%
\section{Correction of Geometric Time Aberrations and Chromaticity}
Let us assume a beam of one species circulating in the ring turn by turn. We observe it in the 
symmetry plane of the ring where the phase-space ellipse is upright (Twiss parameter $\alpha=0$)
and this condition is restored after each turn.
For simplicity we inspect only the horizontal plane, the arguments for the vertical plane are 
the same. The transfer matrix for the full circumference can be expressed by 
the betatron function $\beta$ and the phase advance $\mu$:
\begin{equation}
\label{transfermatrix}
 M =
\begin{pmatrix}
\cos\mu&\beta\sin\mu\\
-\frac{1}{\beta}\sin\mu&\cos\mu\\
\end{pmatrix}
\equiv
\begin{pmatrix}
(x|x)&(x|a)\\
(a|x)&(a|a)\\
\end{pmatrix},
\end{equation}

The matrix elements ($t$$\mid$$xx$), ($t$$\mid$$xa$) and ($t$$\mid$$aa$) can be expressed 
for an achromatic ring as~\cite{yavor,Wolln}:
\begin{eqnarray}
&(t|xx)=\lambda\cdot[(x|x)(a|x\delta)-(a|x)(x|x\delta)] \,,\label{txx}\\
&(t|xa)=\lambda\cdot[(x|x)(a|a\delta)-(a|x)(x|a\delta)] \,,\label{txa}\\
&(t|aa)=\lambda\cdot[(x|a)(a|a\delta)-(a|a)(x|a\delta)] \,,\label{taa}
\end{eqnarray}
where $\lambda$ is a constant with the dimension of inverse velocity.
The coefficients in Eqs.~(\ref{txx}-\ref{taa}) correspond to the terms of the
transfer matrix $M$.

The terms $(x|a\delta)$ and $(a|x\delta)$ are connected via~\cite{Wolln}:
  \begin{equation}
  \label{xad}
  (x|a\delta) \, (a|x) \, = \,(x|x)\,(a|a\delta)\,-\,(a|x\delta)\,(x|a)\,+\,(x|x\delta)\,(a|a).
  \end{equation}
Combining Eqs.~\ref{txx},~\ref{taa}, \ref{xad} we obtain:
  \begin{equation}
  \label{taa_txx1} (t|aa)\,+\beta^2\,(t|xx)\,=\,\frac{\lambda\beta}{sin\mu}[(x|x\delta)\,+(a|a\delta)].
  \end{equation}

The shift of the tune ($Q_{0x}$) of the ring in relative units for a certain momentum shift $\delta$
is called relative chromaticity ($\xi_{1x}$). It can be written as (see appendix):
  \begin{equation}
  \label{naturalchrom_pr} 
  \xi_{1x} =-\frac{1}{4\pi Q_{0x} sin\mu}\cdot[(x|x\delta)+(a|a\delta)].
  \end{equation}
Therefore, by combining Eqs.~(\ref{taa_txx1},~\ref{naturalchrom_pr}) one obtains:
  \begin{equation}
  \label{taa_txx}
  (t|aa)\,+\beta^2\,(t|xx) \,=\, -4\pi\lambda\beta Q_{0x}\xi_{1x}.
  \end{equation}
The emittance of the beam is 
  \begin{equation}
  \varepsilon_x \,=\, x_0 ~ a_0~,
  \end{equation}
and then Eq.\ref{taa_txx} can be written as:
  \begin{equation}
  \label{taa_txx0-rel}
  (t|aa)\,a_0^{2}+(t|xx)\,x_0^{2} \,=\, -4\pi \lambda\varepsilon_{x} Q_{0x}\xi_{1x}~.
\end{equation} 
As one can see the remaining time error for an ion starting in both position and angle 
exactly with the coordinates like the half axis of a matched ellipse ($x_0$ and $a_0$) is still 
proportional to the emittance but also proportional to the chromaticity which can be corrected 
to zero ($\xi_{1x} = 0$).

Next we want to show that for all other starting points in phase space at least the relative 
time deviation over many turns becomes zero.
In general, in a first order achromatic ring with mirror symmetry the two chromatic matrix 
coefficients coincide, $(x|x\delta) \,=\, (a|a\delta)$~\cite{xxd_aad}.
This means the chromaticity correction directly demands, see Eq.\ref{naturalchrom_pr}, 
  \begin{equation}
  (a|a\delta)\,=\,(x|x\delta)=0~.
  \end{equation}
The time deviation by the mixed term $(t|xa)_c$ converges to zero for large number of turns,
  \begin{equation}
  \lim_{n \to \infty}\frac{(t|xa)}{nT_0}\,=\,\lim_{n \to \infty}(t|xa)_c \,=\,0~.
  \end{equation}
When inserting this into the symplectic relation (Eq.\ref{txa}), one sees that also
  \begin{equation}
  \lim_{n \to \infty}\frac{(x|a\delta)}{nT_0}\,=\,0~,
  \end{equation}
and by inserting this into Eq.\ref{taa} also
  \begin{equation}
  \lim_{n \to \infty} \frac{(t|aa)}{nT_0}\,=\, \lim_{n \to \infty} (t|aa)_c\,=\,0~.
  \end{equation}
From this immediately follows also
  \begin{equation}
  \lim_{n \to \infty}\frac{(t|xx)}{nT_0}\,=\, \lim_{n \to \infty} (t|xx)_c\,=\,0~.
  \end{equation}

This means the correction of chromaticity corresponds directly to the correction of the 
otherwise non-vanishing time deviations.
For this task only one condition has to be fulfilled and one family of sextupole magnets is needed.
Together with the same condition for the vertical plane, $(y|y\delta)=0$, and the necessary 
$(t|\delta\delta)=0$, this requires in total three independent sextupole correctors.
Then we can reach a regime where all time deviations vanish in the limit of many turns 
and the isochronous ring behaves like a second-order achromatic system in relative time deviation.


\section{Further Correction of Dispersion}
While the first-order achromaticity of the ring is a basic prerequisite, there are also higher
order contributions to dispersion. 
The mixed aberrations $(t|x\delta)_{c}x\delta$ and $(t|a\delta)_{c}a\delta$ do vanish in the limit 
of many turns, but still for a short observation time one wants to keep their magnitude small.
Again we establish a relationship with the corresponding chromatic aberrations in transverse space.
The symplectic relations for the achromatic ring yield~\cite{yavor,Wolln}:
  \begin{eqnarray}
  \label{txd+tadrealtion}
  &(t|x\delta)=\lambda\cdot[(x|x)(a|\delta\delta)-(a|x)(x|\delta\delta)] \,,\\
  &(t|a\delta)=\lambda\cdot[(x|a)(a|\delta\delta)-(a|a)(x|\delta\delta)] \,,
  \end{eqnarray}
and \cite{xxd_aad} 
  \begin{eqnarray}
  \label{xdd-add}
  (a|\delta\delta)(x|a)-(x|\delta\delta) \, [1-(x|x)]=0
  \label{sympl-bela}
  \end{eqnarray}
In a ring with non-zero phase advance the first order coefficients in Eq.\ref{sympl-bela} cannot 
be zero, and the only solution is
  \begin{eqnarray}
  (x|\delta\delta)\,=\,(a|\delta\delta)\,=\,0~,
  \end{eqnarray}
which then over a single turn leads directly to
  \begin{eqnarray}
  \label{txd+tad}
  &(t|x\delta)\,=\,0~,\\
  &(t|a\delta)\,=\,0~.
  \end{eqnarray}

\begin{figure*}[h]
\centering
\includegraphics[width=110mm]{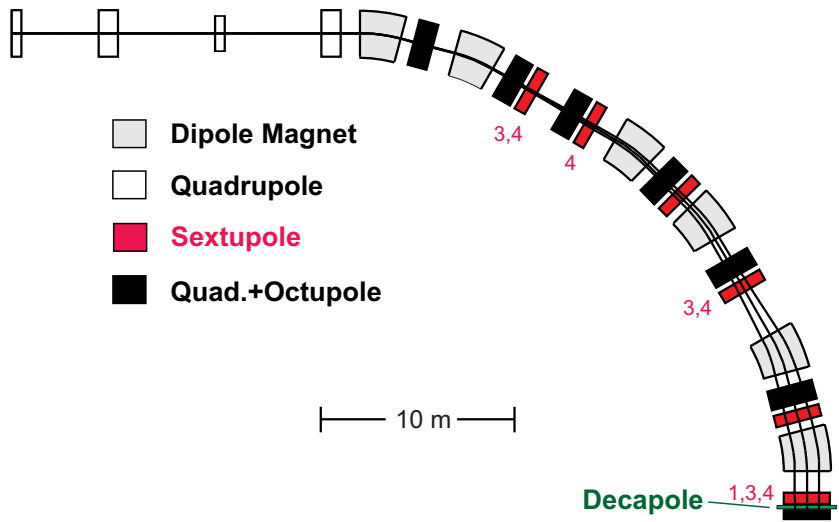}
\caption{Layout of one quarter of the CR. The different magnet types are indicated and the 
numbers mark the sextupole magnets which are used in the correction schemes with only one, 
three or four sextupoles. Inside the apertures dispersion curves are shown.
In the full ring with four-fold symmetry each multipole magnet belongs to a family of magnets 
with same value.
}
\label{cr-quarter}
\end{figure*}

This can be achieved with one additional sextupole which corrects the nonlinearity of dispersion 
after one turn. The influence will be demonstrated in section 6 by a Monte-Carlo simulation.
Fig.\ref{cr-quarter} gives an overview on one quarter of the CR. The positions of sextupoles 
used in different correction schemes are indicated, as well as additional superimposed octupole 
coils inside some of the quadrupole magnets.

We have seen that after one turn the ring must be achromatic at best also to higher orders.
But on the other hand in the arcs we need a well defined dispersion. This must even be
a bit non-linear because the Lorentz factor $\gamma$ is not constant. 
It does not even depend linearly on $\delta$, 
and we want the condition $\gamma=\gamma_t$ (see Eq.\ref{eq1}) to be fulfilled over the whole 
momentum acceptance. The nonlinear dispersion then defines the small but necessary term 
$(t|\delta\delta)$.
In practice this demands symmetric sextupoles for correction to first create and to adjust 
dispersion in the arcs and then to compensate it again.

In the non-dispersive vertical direction the chromatic terms described above do not exist in a 
second-order approximation.
What remains are the purely geometric terms not in diagonal direction inside the phase space ellipse,
$(t|xa)$ and $(t|yb)$, for which also a relation to the chromatic aberrations in positions exists, 
see Eq.\ref{txa}. But as we cannot have directly a full second-order achromat, they are only 
corrected in the limit of many revolutions.

\section{Field Quality}
\subsection{Dipole Homogeneity}

The dipole field homogeneity is crucial for the time resolution.
It can be described by a series of higher order field indices $n_i$
\begin{eqnarray}
\label{dipoleinhom} 
\frac{B(x)}{B_0} = 1 \,- n_1 \frac{x}{\rho_0} \,- n_2 (\frac{x}{\rho_0})^2 \,- n_3 \, 
(\frac{x}{\rho_0})^3 \,- n_4 (\frac{x}{\rho_0})^4 \,- n_5 (\frac{x}{\rho_0})^5 \\
\text{with }~n_1=0,~n2=1.15,~n_3=0.23,~n_4=568,~n_5=126~.
\end{eqnarray}
The sector radius $\rho_0$ is $8.125\,m$ and
the given values refer to the case of maximum field strength ($B=1.6\,T$).
The chromaticity depending only on the quadrupole fields in the ring, usually called 
"natural chromaticity", is one contribution, but
the magnet design predicts a large second order deviation, which causes an effect even stronger 
than the natural chromaticity, making chromaticity worse in horizontal 
direction and overcompensating it in vertical direction.
But as $n_2$ represents a pure sextupole component it can be completely corrected with three 
sextupoles including the $(t|\delta\delta)$ term and with no increase in maximal sextupole 
magnet strength compared to the case of a homogeneous dipole field.

By exciting one assumed multipole component of order $i$ in the main field of all dipoles one sees 
that each multipole component leads to a corresponding time error $(t|\delta ^i)$ after one turn.
These deviations add up over many turns and do not average out. 
Compensation must be made with adjustable multipoles. So far sextupoles and octupoles superimposed 
to some of the quadrupoles and one decapole are foreseen, see Fig.\ref{cr-quarter}. 

\begin{figure*}[tb]
\centering
\includegraphics[width=\textwidth]{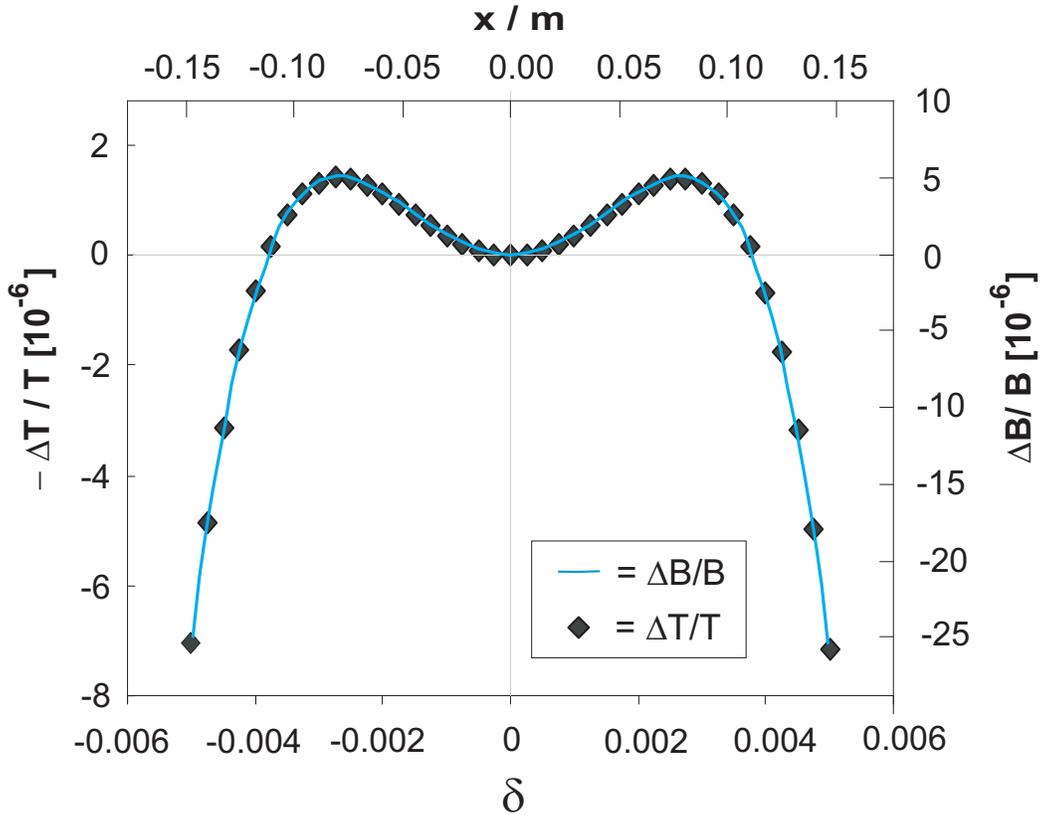}
\caption{
Comparison of dipole inhomogeneity from Eq.\ref{dipoleinhom} as function of position $x$ with the resulting 
time deviation as function of $\delta$. For comparison the curves are shifted on top of each other 
with the help of an effective dispersion coefficient $(x|\delta)_\text{eff}=29\,m$.
}
\label{B-tof-shape}
\end{figure*}

Fig.\ref{B-tof-shape} shows the remaining field shape deviations of order four and higher. 
For demonstration the $n_2$ component in the dipole was assumed to be compensated 
so that only higher order terms (mainly $n_4$) contribute.
Sextupoles were adjusted to overcompensate the second-order dependence
to minimize the overall field deviation. 
Still the time spread assuming a momentum distribution according to the full CR acceptance 
would be limited to $\sigma_T/ T \approx 10^{-6}$.
Clearly, a weak decapole field close to the maximum of dispersion makes compensation 
more effective and easier to adjust.

In a simplified model the relation between the relative field quality $\Delta B/B(x)$ and 
the time deviations $\Delta T/T(\delta)$ can be described with an effective dispersion of 
$(x|\delta)_\text{eff}=29\,m$. Then the ratio becomes
\begin{equation}
\label{fieldhom} 
\frac{\Delta T(\delta)~/T_0}{\Delta B((x|\delta)_\text{eff} \,\delta )~ /B_0} \, \approx \, -0.28 .
\end{equation}
This means the required field homogeneity must be very close to the desired mass resolution.
As a dipole magnet with large gap will not provide this many multipole correctors must be foreseen.

\subsection{Fringe Fields}
The fringe fields of the magnets include also many more multipole terms and the CR is 
designed for large apertures. Therefore, the influence of the fringe fields was studied in detail.
The analysis is based on a realistic fringe field distribution as calculated for the magnet 
design of the CR dipole and quadrupole magnets \cite{Gorda}.
They were approximated by Enge functions and corresponding transfer matrices were calculated
using the program of ref.\cite{Har90}. An alternative is the method of fringe field integrals
\cite{Mat70,Mat72} which yields practically the same results. Both are implemented into the 
GICOSY code \cite{GICOSY,GICOSY2} used for the analysis.

The simple quasi-analytic fringe field integral equations reveal that the first
influence of the quadrupole fringe fields lies in a change of position and angle 
when describing the effect by a homogeneous main field with an additional short fringe field
matrix.
Each entrance and exit matrix contains terms $(x|x)=-(a|a)=-(y|y)=(b|b)= 1 \pm w $
with $w$ being proportional to the gradient and aperture radius of the magnet. In case of 
the large quadrupoles of the CR with aperture radius $r_0=0.168\,m$ and the strongest excitation 
($B=0.4898\,T$ for $B\rho=13\,Tm$) the value can reach $w=0.000727$.
Over the whole ring this still leads to a matched beam with slightly altered beam parameters 
and only small dispersion mismatch. 
When including them the tunes of $Q_x=2.153$ and $Q_y=3.186$ shift by $-3.7\%$ or $-1.2\%$, respectively.
Even without a readjustment of the quadrupole setting the compensation for different velocities 
$(t|\delta)$ still stays on the level 
of few times $10^{-6}$. Second order terms are not significantly affected, only slightly by the natural 
chromaticity of the changed first order focusing, see table\ref{tuneshifttable}.

However, the well known third order aperture aberrations of the quadrupole fringe fields make 
a significant difference.
After the described compensation of the second-order path length differences and counterbalancing
of dipole inhomogeneities with other multipoles, which is largely possible for the given numbers,
they introduce the strongest limitation of ToF resolution.
The calculations show that this is largely independent of the gap size and fringe field shape.
The main contributions exist even for small aperture radius with a very short extension of the
fringe field as long as the shape follows a realistic shape and not a simple cut off which
actually violates Maxwell equations.

The dipole magnets in CR have no edge angle for additional focusing.
By the nature of the fringe fields some additional focusing in first order in vertical 
direction is added \cite{Mat70}. But its magnitude is smaller than in the quadrupole fringe fields.
However, in any magnetic sector field the fringe field introduces an additional focusing 
in vertical direction dependent on the horizontal beam direction. Its size is practically 
constant as in leading order it directly depends only on the deflection radius, 
$(b|ya)=1/\rho_0 $ \cite{Mat70}. This couples the vertical focusing to the horizontal dispersion
and changes the vertical chromaticity.

In addition the vertical focusing itself is also associated with chromatic terms depending 
on the shape of the field distribution.
When adding realistic dipole fringe fields to a CR with adjusted first order tunes 
they shift the tune in vertical direction by up to
  \begin{equation}
  \Delta Q_y \,=\, \xi_y \, Q_{0y} \, \delta_{max}~.
  \end{equation} 
This shift is small compared to the natural chromaticity, see table[\ref{tuneshifttable}], and
the tune in horizontal direction is practically not affected. 
The sign of the chromaticity from the dipole fringe fields even compensates partially the natural 
chromaticity. All which is required is a smaller readjustment of the sextupole setting.

\begin{center}
\begin{table}
\caption{Chromatic tune shifts for $\delta=0.5\%$ by different influences separately. 
The shifts are given relative to a ring with no chromaticity.}
\begin{tabular}{l | c | r }
\label{tuneshifttable}
                                       & $\Delta Q_x$& $\Delta Q_y$ \\
\hline
natural chromaticity (sext. off)        & -0.0196   & -0.0138 \\
dipole field with n$_2$=1.15            & -0.0260   &  0.0176 \\
dipole fringe fields                    & $<\,10^{-4}$ &  0.0032 \\
quadrupole fringe fields                & $<\,10^{-4}$ &  $<\,10^{-4}$ 
\end{tabular}
\end{table}
\end{center}

In general most sensitive to higher order multipole components are the terms for 
isochronicity depending purely on $\delta$. 
This means after each adjustment of chromaticity to reduce the transverse path 
length differences one sextupole, octupole and decapole must be adjusted.
The size of the other remaining chromatic transverse path length differences in third or 
higher order can amount to $\Delta T/T \approx 10^{-6}$ depending on the exact multipole
correction.


\section{Results of Numerical Simulations}

Numerical simulations were done by tracing single ions directly inside the GICOSY
program by transforming their coordinates by up to fifth order transfer matrices for 
each optical element. This method differs from inserting the initial coordinates in one overall 
transfer matrix by not truncating higher orders which appear in a 
multiplication of lower order matrices with order larger than one. 
In the simulations an initial five-dimensional phase space area was assumed with an emittance 
of $100$ mm mrad in each plane and a momentum spread of $\pm\,0.5\%$.
The shape of the initial ellipses was chosen to have matched beams in x and y plane. The ellipses 
in both planes were filled independently.
The stepwise tracing of ions also makes insertion of collimators possible, however for the
chosen size of phase space the transmission is larger than $98\%$ and in the results shown
no collimators were used.
This approach is similar to the program MOCADI \cite{Iwa2011} which was used as a cross check up 
to third order.

\begin{figure*}[ht]
\centering
\includegraphics[width=\textwidth]{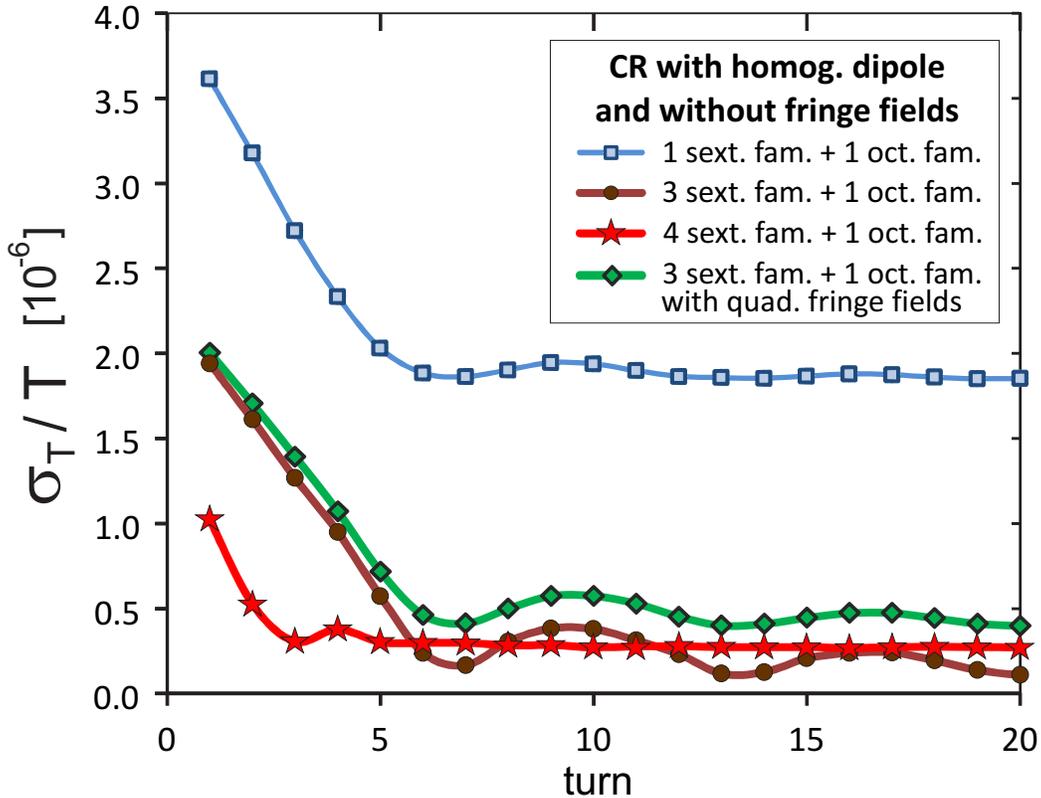}
\caption{Standard deviation of relative ToF as function of turns. The Monte-Carlo simulations show
three cases:  one sextupole to correct only $(t|\delta\delta)$, three sextupoles to correct in addition 
chromaticity in x and y plane, and 4 sextupoles to correct as well second order dispersion
$(x|\delta\delta)$. In one calculation (green curve) fringe fields on the quadrupoles 
were considered in addition.
}
\label{mocadi-turns}
\end{figure*}

The averaging of time aberrations over the full phase space allows a realistic comparison of the 
time resolution.
To better understand the resolution over a lower number of turns Fig.\ref{mocadi-turns} shows 
the standard deviation in time as function of turn number in a simplified ring with perfectly 
homogeneous dipole magnets and sharp cut off multipole fields.
As one can see, convergence to an almost constant limit goes fast, which justifies our approach 
of correcting mainly terms not vanishing over many turns.
When correcting only the velocity dependence the time resolution for large emittance remains 
still too large for good mass measurements of heavy ions. The chromaticity correction improves 
this limit by almost a factor of ten. With three sextupoles there are small oscillations in the 
time spread which result from non-matched dispersion in higher order with the corresponding time 
errors, see Eq.\ref{txd+tad}. The size of this dispersion oscillates with the frequency 
of the horizontal tune. A fourth sextupole can correct this and thereby make convergence faster. 
However, more and stronger sextupoles are needed which inflict more higher order aberrations.
What remains after many turns are time errors mainly of higher order, even though 
the simulations already include one octupole family located near the maximum of dispersion 
in the arc of the ring. It is needed to correct the largest third-order contributions to the 
revolution time, which is the purely chromatic term $(t|\delta\delta\delta)$.

\begin{figure*}[h]
\centering
\includegraphics[width=\textwidth]{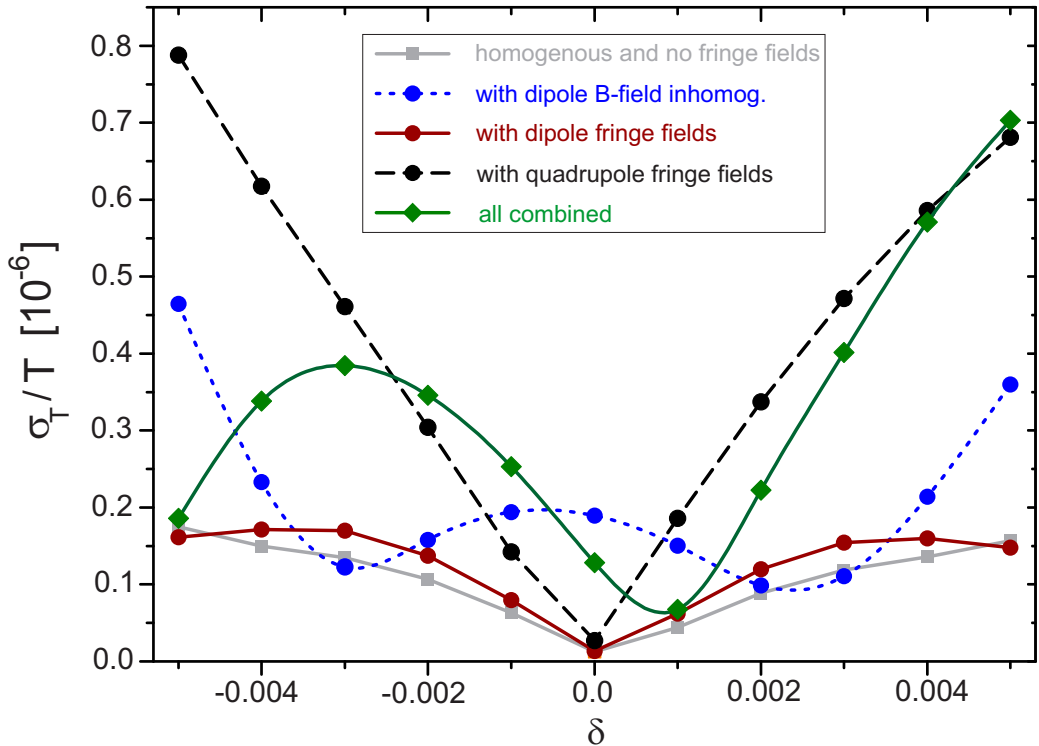}
\caption{Relative time deviation after 100 turns for the full accepted transverse phase space 
as function of momentum deviation $\delta$.
After correction of pure $\delta$ dependence and chromaticity with sextupoles 
the remaining influence of including dipole inhomogeneities and quadrupole or dipole fringe fields 
is shown. The green curve shows all effects combined.}
\label{dt-effects}
\end{figure*}

The expected imperfections of the dipole magnets require stronger sextupole corrections
than without and also a readjustment of at least one octupole for the condition $(t|\delta\delta\delta)=0$.
Then the decapole component of the dipole dominates. 
Already Fig.\ref{B-tof-shape} makes it clear that without adding one separate weak decapole 
the goal in resolution cannot be reached.
The remaining ToF deviations for different field imperfections are shown in Fig.\ref{dt-effects}
by the standard deviation in ToF as function of momentum deviation.
In these calculations a correction of chromaticity with sextupoles and a correction of the pure 
momentum dependence up to fourth order was considered with the help of three sextupoles, one octupole 
and one decapole.
As fringe fields with real extension were inserted also a small readjustment
of quadrupole and sextupoles was required.
The influence of the additional multipoles in the dipole magnet can mostly be corrected but in 
combination with the large emittance some contributions remain which depend on the actual sextupole 
setting. Here values for the simple case with only three sextupoles are shown.
Also the fringe fields limit the ToF resolution. The influence of the quadrupole fringe fields dominates.
In general limitations from higher order mixed components remain. The dependence on transverse
motion alone is corrected very well by sextupoles but in combination with dispersion more
deviations show up.

That quadrupole fringe fields have a basic unavoidable influence is demonstrated by
adding them to the perfect pure multipole fields in Fig.\ref{mocadi-turns}. 
To show the basic influence the pole tip radius size was reduced very much so that the fringe field 
extension becomes very short while keeping the gradient constant. 
In this case the first-order focusing becomes identical to the case without fringe fields. 
However, in third order many aberrations remain which do not vanish for a real field distribution 
and are even present for very small gap size.
As apertures were not used in the calculation the CR with small quadrupole radius still has 
the same acceptance.
One can see that even then the quadrupole fringe field represents a 
major contribution to the limit in ToF resolution.

\begin{figure*}[h]
\centering
\includegraphics[width=\textwidth]{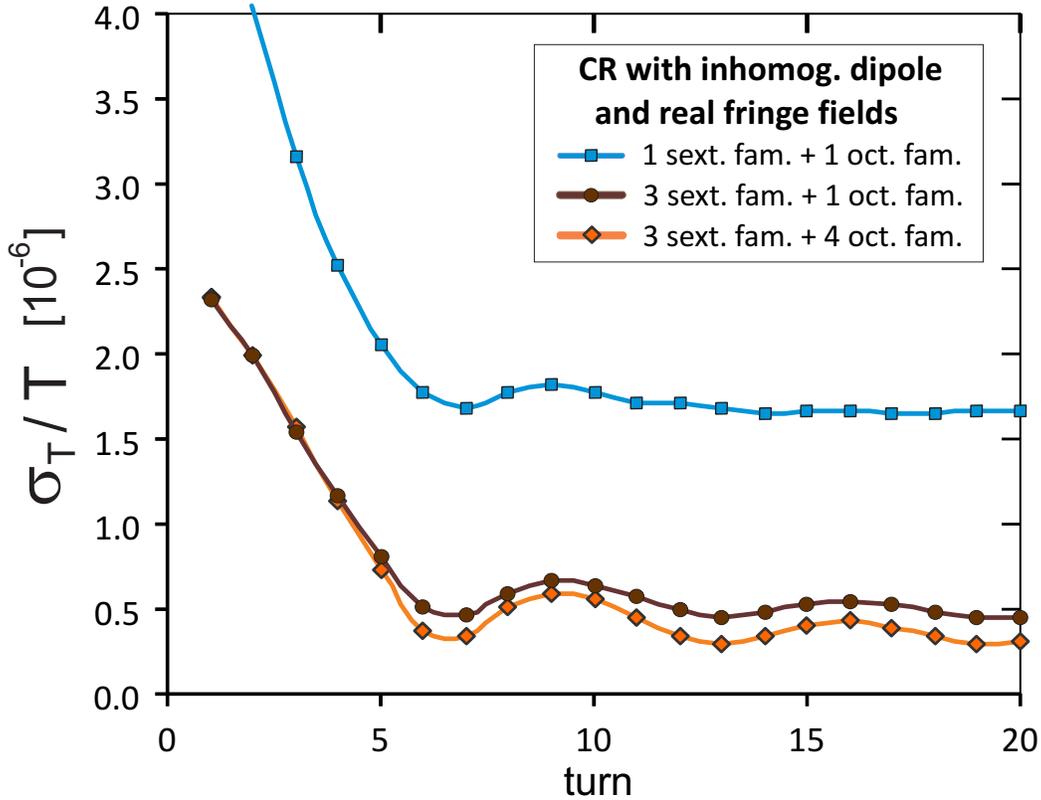}
\caption{Time spread as function of turn number, similar to Fig.\ref{mocadi-turns} 
but with realistic dipole inhomogeneities and fringe fields. In addition to only
one octupole also four octupoles families were used for correction.}
\label{mocadi-turns2}
\end{figure*}

Considering all influences the dependence on turn number is shown again in Fig.\ref{mocadi-turns2},
now with all fringe fields and dipole inhomogeneities.
Applying in addition further octupole corrections also some of the third order ToF deviations
can be corrected but the effect of quadrupole fringe fields is more difficult to compensate.
Contrary to the dipole inhomogeneity it does not correspond to a simple multipole
component which can be compensated with an opposite field nearby. 
More terms coupling the longitudinal and transverse motion become important.
With additional four weak superimposed octupoles the result is improved a bit, 
see Fig.~\ref{mocadi-turns2}.
One can reach a resolution of up to $\sigma_T\slash{T}\approx 3.0 \cdot 10^{-7}$, which 
corresponds to a mass resolving power of $m/\sigma_m =1.2 \cdot 10^{6}$ or $\sigma_m=100\,keV$ 
at $m=130\,u$. 
Further improvement in mass precision can be reached by detecting many ions of one species.

\section{Calibration and Tuning}

In principle ions with different $m/q$ can be separated if their mean time separation 
$\Delta T$ is larger than the full time spread of the beam over a number of turns and
including detector resolution.
However, only ions of very similar $m/q$ can be in a good velocity region to be isochronous.
For other $m/q$ but same magnetic rigidity $B\rho$ the velocity $\gamma$ must be different
from $\gamma_t$.
In practice this leads to the problem of finding enough calibrants with well known masses.
Assuming perfect isochronicity for an ion of $m/q=140/50\,$u/e leads for $m/q=139/50\,$u/e to 
a large time spread of $\sigma_T/T=1.1\cdot 10^{-5}$  in case of full 
acceptance of the CR with $\sigma_{\delta}=0.003$.

At the same time a method for tuning of isochronicity with all the mentioned corrections must 
be found. Scanning over the momentum acceptance with a narrow emittance beam of variable energy
is used in the ESR \cite{Hau2000}, but there is no electron cooler foreseen in the CR and
stochastic cooling will not work in the isochronous mode.
Instead a second ToF detector on the straight section of the CR shall be installed \cite{ILIMA-TP}.
From the time difference between the two detectors the velocity can be deduced,
whereas over a full turn the revolution time does not depend on velocity. 
From the velocity the magnetic rigidity can be calculated and
by following many ions of different velocity a full isochronicity curve can be recorded.
In addition with the help of slits and steerers in the preceding Super-FRS separator \cite{sfrs} 
and transfer beamline also regions of the transverse phase space can be selected.

The required resolution in time-of-flight for this velocity measurement is not as high as for
the exact mass determination. It is sufficient to provide enough calibration points in $\delta$
over the whole CR momentum acceptance and to be able to distinguish ions of different 
mass number. Even with masses only from theoretical predictions the uncertainty in the conversion 
from velocity to $B\rho$ and $\delta$ is small enough.

At best the detectors are located on two ends of a straight section. The distance in case
of the CR is $22\,m$. The time deviation of the present ToF detector was determined experimentally
and amounts to $\sigma_t=37\,ps$ \cite{Kuz2011}. This leads to a time resolution 
of $\sigma_t /t=6 \cdot 10^{-4}$, which is improved by observing an ion over many turns. 
Even when considering the limited efficiency of the detector, which will not give a valid time signal 
for each turn, this will still yield $\Delta t/t > 3 \cdot 10^{-5}$ after 50 revolutions.
At $\gamma=1.67$ this corresponds to an error in $\delta$ of $7 \cdot 10^{-5}$. 
Compared to the ring acceptance of $\pm 5 \cdot 10^{-3}$ this gives a good resolution.
On this level the described second order path length differences which still exist over the 
straight section are not critical.

It is noteworthy that the energy loss of $^{140}$Sn ions in two foils of $20 \, \mu g/cm^2$ 
over 50 turns corresponds to $\delta = -8.6 \cdot 10^{-5}$. 
That means it is possible to trace the energy loss in the foils. One could even evaluate the 
revolution time only in a narrowed region that corresponds to good isochronicity. 
By this $B\rho$ tagging also ions with larger difference in $m/q$ could be included in the 
analysis. One would simply evaluate them only when they are close to $\delta=0$. By 
the energy loss in the foil even ions with initially too high $B\rho$ will pass this point if the 
difference is less than the CR acceptance.
This would lead to a large gain in acceptance compared to $B\rho$ tagging as performed
today by selection with slits inside the FRS or ESR \cite{Gei2006}.

\section{Conclusion}
We have presented a method for correcting second order time-of-flight aberrations in a storage ring.
As a full correction over one turn is not feasible, instead a correction over many turns is proposed.
The necessary requirements on the higher-order multipole correctors were derived.
A Monte-Carlo simulation for the full phase-space acceptance shows
that the main part of the not corrected terms average out fast.
These correction must take into account the unavoidable influence of dipole inhomogeneities and 
of extended fringing fields which were investigated in detail for the CR.
Other multipole components can be compensated with more multipole correctors but
especially the spherical aberrations in the fringe field of quadrupoles are hard to compensate.
Nevertheless, the achievable time resolution is sufficient for mass measurements even for the full
acceptance of the CR. 
A detection scheme with two detectors in the ring helps in tuning
and by tagging properties of individual ions the accuracy can be improved.

\section{Appendix}
Here we derive the expression for the nonlinear chromaticity described in matrix coefficients.
The phase advance $\mu$ can be expanded with respect to $\delta$ as~\cite{Tak04}:
\begin{equation}
\label{musumm} 
\mu = \sum_{n=0}^{\infty} \mu_{n}\delta^{n},\quad \text{or}
\quad cos \, \mu = \sum_{n=0}^{\infty} \chi_{n}\delta^{n}.
\end{equation}
Since $\cos\mu$ is the trace of the transfer matrix $M$ (see Eq.~\ref{transfermatrix}), 
which can also be expanded with respect to $\delta$ one obtains 
\begin{equation}
\label{chitr} 
\chi_{n} = \frac{1}{2} \, Tr(M_{n}),
\end{equation}
where $M_{n}$ is in terms of the transfer matrix coefficients:
\begin{equation}
\label{matrixmn} M_{n} = \left(
\begin{array}{cc}
(x|x\delta^{n}) & (x|a\delta^{n})  \\
(a|x\delta^{n}) & (a|a\delta^{n}) \\
\end{array} \right),~n=0,~1,~2,~3,..
\end{equation}
Thus, from Eqs.(\ref{musumm},\ref{matrixmn}) one can extract the phase advance described 
in matrix coefficients and correspondingly obtain the chromaticity:
\begin{equation}
\label{chromsumm} 
\xi \,=\, \frac{1}{\delta} \, \frac{\Delta Q}{Q_{0}} 
\,=\, \frac{\Delta\mu}{2\pi\delta Q_{0}} 
\,=\, \frac{1}{2\pi Q_{0}}\sum_{n=1}^{\infty} \mu_{n}\delta^{n-1}~.
\end{equation}

\section*{Acknowledgment}
We are thankful to the colleagues from the storage ring departments at GSI and the 
ILIMA collaboration.
This work was carried out with support from German DAAD and Serbia in the PP 
program "Rings for NUSTAR", nr. 50752297.

%


\begin{thebibliography}{99} 
%
\bibitem{cr}
A. Dolinskii, K. Beckert, P. Beller, B. Franzke, F. Nolden, M. Steck,
EPAC Proceedings, 572-574 (2002).
%
\bibitem{fair}
FAIR Baseline Technical Report, GSI Darmstadt (2006).
\texttt{http://www.fair-center.de}
%
\bibitem{sfrs}  
H. Geissel et. al., Nucl. Instr. and Meth. in Phys. Res. B 204 (2003) 71-85.
%
\bibitem{Hau2000}
M. Hausmann et. al., Nucl. Instr. and Meth. in Phys. Res. A 446 (2000) 569-580.
%
\bibitem{Tu2011}  
X.L. Tu et al., Nucl. Instr. and Meth. in Phys. Res. A 654 (2011) 213.
%
\bibitem{Yam2008} 
Y. Yamaguchi et al., Nucl. Instr. and Meth. in Phys. Res. B 266 (2008) 4575.
%
\bibitem{Dol2007}   
A. Dolinskii, S. Litvinov, M. Steck, H. Weick,
Nucl. Instr. and Meth. in Phys. Res. A 574 (2007) 207-212.
%
\bibitem{Dol2008}  
A. Dolinskii, H. Geissel, S. Litvinov, F. Nolden, M. Steck, H. Weick, 
Nucl. Instr. and Meth. in Phys. Res. B 266 (2008) 4579.
%
\bibitem{stori11}
S. Litvinov et. al., "Status of the Storage Ring Design at FAIR", 
PoS(STORI11)026, Frascati, Italy (2011),
%
\bibitem{Wol90}  
H. Wollnik, Nucl. Instr. and Meth. in Phys. Res. A298 (1990) 156.
%
\bibitem{Cou71}  
E.D. Courant, Particle Accelerators, 2, 117-119 (1971).
%
\bibitem{Haw09}  
P.W. Hawkes, Phil. Trans. R. Soc. A 367 (2009) 3637-3664.
%
\bibitem{Bro79}  
K.L. Brown, SLAC-Pub-2257 (1979), and
K.L. Brown, IEEE Trans. Nucl. Sci. NS-26 3, 3490 (1979).
%
\bibitem{Wou85}   
J.M. Wouters, D.J. Viera, H. Wollnik, H.A. Enge, S. Kowalski, K.L. Brown, 
Nucl. Instr. and Meth. in Phys. Res. A 240 (1985) 77-90.
%
\bibitem{Wol87b}  
H. Wollnik, Nucl. Instr. and Meth in Phys. Res. B 26 (1987) 267.
%
\bibitem{Wan96} 
Weishi Wan and Martin Berz, Physical Review E 54, 2870 (1996).
%
\bibitem{Dra87} 
A.J. Dragt, Nucl. Instr. and Meth in Phys. Res. A 258 (1987) 339.
%
\bibitem{Ser87}  
R.V. Servranckx, Nucl. Instr. and Meth. in Phys. Res. A 258 (1987) 525-535.
%
%
%
\bibitem{carey}
D.C. Carey, "The optics of charged particles beams",
Harwood Academic Publisher, New York 1987.
%
\bibitem{yavor}
M. Yavor, "Optics of Charged Particle Analyzers",
Advances in Imaging and Electron Physics vol.~157, Elsevier Inc. 2009.
%
\bibitem{Wolln}
H. Wollnik, "Optics of charged particles", Academic Press, Florida 1987.
%
\bibitem{xxd_aad}
B. Erdelyi, J. Maloney, J. Nolen, 
Physical Review Special Topics~-~Accelerators and Beams 10, 064002 (2007).
%
%
\bibitem{Gorda}
Oleksii Gorda, PhD thesis, University of Frankfurt (2011).
\texttt{\small http://www-alt.gsi.de/documents/DOC-2011-Aug-28-1.pdf}
%
\bibitem{Har90}
B. Hartmann, Nucl. Instr. and Meth. in Phys. Res. A297 (1990) 343-353.
%
\bibitem{Mat70} 
H. Matsuda, H. Wollnik, Nucl. Instr. and Meth. in Phys. Res. 77 (1970) 40.
%
\bibitem{Mat72} 
H. Matsuda, H. Wollnik, Nucl. Instr. and Meth. in Phys. Res. 103 (1972) 117.
%
\bibitem{GICOSY}
H. Wollnik, B. Hartmann and M. Berz, AIP Conf. Proceedings, 177, 74-85 (1988).
%
\bibitem{GICOSY2}
M. Berz, H.C. Hoffmann and H. Wollnik, Nucl. Instr. and Meth. in Phys. Res. A258 (1987) 402. \\
and \texttt{\small http://web-docs.gsi.de/$\sim$weick/gicosy}
%
\bibitem{Iwa2011}
N. Iwasa, H. Weick, H. Geissel, Nucl. Instr. and Meth. in Phys. Res. B 269 (2011) 752.\\
\texttt{\small http://web-docs.gsi.de/$\sim$weick/mocadi}
%
\bibitem{ILIMA-TP} 
ILIMA collaboration, technical proposal (2005). \\
\texttt{\small http://www.fair-center.de/fileadmin/fair/experiments/}
\texttt{\small NUSTAR/Pdf/ILIMA\_tp\_20050120-printed.pdf}
%
\bibitem{Kuz2011}
Natalia Kuzminchuk, PhD thesis, University of Giessen (2011).
\texttt{\small http://geb.uni-giessen.de/geb/volltexte/2012/8726}
%
\bibitem{Gei2006} 
H. Geissel et al., Hyperfine Interact 173 (2006) 49-54.\\
%
\bibitem{Tak04}
Masaru Takao, Hitoshi Tanaka, Kouichi Soutome, and Jun Schimizu,
Physical Review E 70, 016501 (2004).
%

\end{thebibliography}
\end{document}